\documentclass[english,prb,reprint,twocolumn,superscriptaddress]{revtex4}
\usepackage[LGR,T1]{fontenc}
\usepackage[latin9]{inputenc}
\usepackage{textcomp}
\usepackage{mathrsfs}
\usepackage{amstext}
\usepackage{graphicx}
\usepackage{esint}

\makeatletter

\DeclareRobustCommand{\greektext}{%
  \fontencoding{LGR}\selectfont\def\encodingdefault{LGR}}
\DeclareRobustCommand{\textgreek}[1]{\leavevmode{\greektext #1}}
\ProvideTextCommand{\~}{LGR}[1]{\char126#1}

\@ifundefined{textcolor}{}
{%
 \definecolor{BLACK}{gray}{0}
 \definecolor{WHITE}{gray}{1}
 \definecolor{RED}{rgb}{1,0,0}
 \definecolor{GREEN}{rgb}{0,1,0}
 \definecolor{BLUE}{rgb}{0,0,1}
 \definecolor{CYAN}{cmyk}{1,0,0,0}
 \definecolor{MAGENTA}{cmyk}{0,1,0,0}
 \definecolor{YELLOW}{cmyk}{0,0,1,0}
}

\usepackage[english]{babel}
\def\urlprefix{}
\def\url#1{}
\addto\captionsenglish{%
}

\usepackage[none]{hyphenat}

\usepackage{babel}

\usepackage{babel}

\makeatother

\usepackage{babel}
\begin{document}

\title{Thermally Driven Topology in Frustrated Systems}

\author{Jie-Xiang Yu}

\thanks{Correspond to: jiexiang.yu@unh.edu}

\affiliation{Department of Physics and Materials Science Program, University of
New Hampshire, Durham, New Hampshire 03824, USA}

\author{Morgan Daly}

\affiliation{Department of Physics and Materials Science Program, University of
New Hampshire, Durham, New Hampshire 03824, USA}

\author{Jiadong Zang}

\affiliation{Department of Physics and Materials Science Program, University of
New Hampshire, Durham, New Hampshire 03824, USA}
\begin{abstract}
Non-trivial topology in a two-dimensional frustrated spin system with
the Dzyaloshinskii-Moriya (DM) interaction was investigated by Monto
Carlo simulations. At finite temperatures, thermally driven topology
was discovered and was found to be dominant at low magnetic field.
This topological charge has a quadratic relation with the DM interaction
and linear realtions with the external magnetic field or the uniaxial
magnetic anisotropy. We also proposed a real frustrated system, the
Mn-Bi mono-layer film with exceedingly large DM interaction, to enable
thermally driven topology. Other topological non-trivial phases in
high magnetic field region were also discussed in this real system.
\end{abstract}
\maketitle

\section{Introduction}

Low dimensional magnetic systems have attracted long term interests
due to numerous exotic phenomena therein. Its recent marriage to topology
has stimulated the development of skyrmion physics. The magnetic skyrmion
is a two-dimensional (2D) topological spin texture mainly discovered
in B20 chiral magnets.\citep{Muhlbauer_2009,Neubauer_2009,Bogdanov_1994,Yu_2011_nmat,Jiang_2016,Jiang2017}
It was first observed in MnSi by small angle neutron scattering and
later confirmed in $\mathrm{Fe}_{0.5}\mathrm{Co}_{0.5}\mathrm{Si}$
thin slabs in terms of real space imaging by Lorentz transmission
electron microscopy.\citep{Muhlbauer_2009,Neubauer_2009,Yu_2010_nat}
Although in reality, skyrmions are extensively observed and discussed
in three dimensional (3D) samples, the concept of skyrmion is still
strictly defined in 2D. Skyrmions are uniformly stacked in 3D. Topology
of the configuration itself has induced novel properties in skyrmion
dynamics and electronic/magnonic transports. 

Topology of the skyrmion can be mathematically described in terms
of a topological index dubbed the topological charge $Q_{T}$\citep{Berg_1981,Nagaosa2013,Yin_2016},
which measures the coverage of a spin configuration onto a unit sphere.
Given $\Theta$ and $\Phi$ the polar and azimuthal angles, respectively,
of each spin $\mathbf{S}$, the topological charge is given by 
\begin{equation}
Q_{T}=\frac{1}{4\pi}\int\text{\ensuremath{\sin\Theta}}d\Theta d\Phi.\label{eq:tcharge}
\end{equation}
Geometrically, it simply counts the total solid angle, in units of
$4\pi$, enclosed by the configuration. In a compact manifold, $Q_{T}$
is always an integer number. For a ferromagnetic state, the total
solid angle is zero, so that $Q_{T}=0$, while for a skyrmion, it
completely covers a unit sphere and total solid angle is 4\textgreek{p},
so that $Q_{T}=\pm1$. In a smooth spin configuration, $Q_{T}$ can
be written in the well adopted form $Q_{T}=\frac{1}{4\pi}\int d^{2}r\mathbf{S}\cdot(\partial_{x}\mathbf{S}\times\partial_{y}\mathbf{S})$.
However, its geometrical definition in Eq. \ref{eq:tcharge} is more
universal and applicable even in highly disordered spin configurations. 

Skyrmions are well modeled in 2D chiral magnets, in which broken inversion
symmetry induces the Dzyaloshinskii-Moriya (DM) interaction. In a
finite window of the $B-T$ phase diagram, the skyrmion crystal phase
can be figured out. It is not surprising to have nonzero topological
charge in this topologically nontrivial phase. However, as reported
by authors' recent work, nonzero topological charge exists in a wide
range in the phase diagram. \citep{Hou2017} Particularly with large
magnetic field, the topological charge is nonzero at finite temperatures
and peaks around the melting temperature of the ferromagnetic state
though it approaches to zero at both low and high temperatures limit.
Such emergence of the topological charge is thus driven by thermal
fluctuations. In this regime, the spin configuration is completely
random, and the topological charge spreads out in the whole lattice.
Symmetry consideration suggests that the topological charge is proportional
to the external magnetic field $B$ and quadratic order of the DM
interaction $D^{2}$. This thermally-driven topology can be detected
by thermal magnon Hall effect.\citep{Huang2012,Mochizuki_2014,Iwasaki2014,Hirschberger2015,Lee_2015}
Thus, the correspondence between the skyrmion and topological charge
is not one to one. The skyrmion definitely leads to nonzero topological
charge, but nonzero topological charge does not necessarily indicate
the presence of skyrmions. Nonzero topological charge should exist
in a variety of 2D spin models with DM interactions. 

\begin{figure}
\includegraphics[width=1\columnwidth]{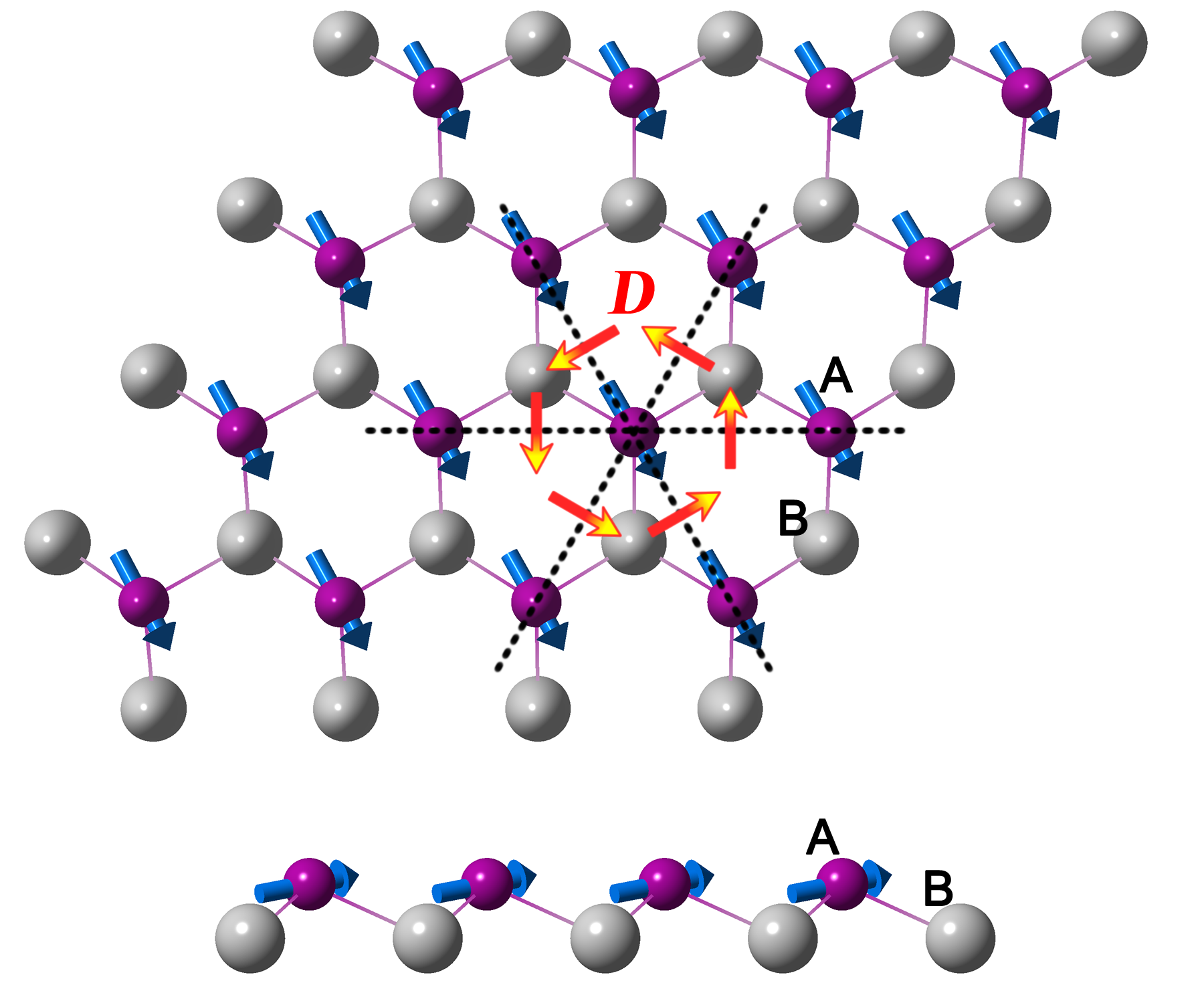}

\caption{Top view and side view for the crystal structure of a 2D hexagonal
lattice with $A-B$ sublattices. $A$ has local magnetic moment. Three
dashed lines give six nearest neighbors for one $A$ site. The direction
of DM interaction $D$ is given by six arrows. }
\label{fig:lattice_AB}
\end{figure}

In 2D chiral magnets, skyrmion phase and thermally driven topological
phase coexist when magnetic field is low so that it is hard to distinguish
them. In this work, we show the presence of pure thermally driven
topological charge in a frustrated antiferromagnetic system. It is
a simple hexagonal lattice, which can be regarded as a 2D hexagonal
boron nitride structure with buckling, shown in Fig. \ref{fig:lattice_AB}.
Two sublattices, labeled by $A$ and $B$, are in different atomic
mono-layers. Magnetic atoms are located at sublattice $A$ and heavy
atoms such as $4d$ or $5d$ transition metal with strong spin-orbit
coupling (SOC) are located at $B$. This system has the point group
of $C_{3v}$ without inversion symmetry and it is also a prototype
of many non-centrosymmetric magnetic monolayer film systems such as
Fe/Ir(111) and Fe/Re(0001).\citep{Heinze2011,Grenz2017,Palacio-Morales2016}
Then the Hamiltonian in this model is given by
\begin{equation}
\begin{array}{ccl}
\mathscr{H} & = & \sum_{<i,j>}\left[J\mathbf{S}_{i}\cdot\mathbf{S}_{j}+\mathbf{D}_{ij}\cdot(\mathbf{S}_{i}\text{\texttimes}\mathbf{S}_{j})\right]\\
 &  & -\sum_{i}\left[\frac{1}{2}K_{u}S_{iz}^{2}+BS_{iz}\right]
\end{array}\label{eq:H_hex}
\end{equation}
where in the first term, $S_{i}$ and $S_{j}$ are the two nearest
neighbor local magnetic moments at sites $i$ and $j$ of sublattice
$A$ respectively. Each site has six nearest neighbors. The Heisenberg
interaction $J$ originates from the superexchange between two neighboring
$A$ sites along $A-B-A$ as well as direct exchange along $A-A$.
The direction $\hat{\mathbf{D}}_{ij}$ of DM interaction $\mathbf{D}_{ij}=D\cdot\hat{\mathbf{D}}_{ij}$
is perpendicular to the bond connecting site $i$ and $j$ according
to the Moriya rule \citep{Moriya1960}, so it is counterclockwise
around one moment, shown in Fig. \ref{fig:lattice_AB}. The second
term includes on-site uniaxial magnetic anisotropy $K_{u}$ and external
magnetic field $B$ along $z$ direction. 

Antiferromagnetic coupling, namely $J>0$, in this simple hexagonal
lattice leads to a spin frustrated system. It has been reported that
the skyrmion phase is, in principle, possible in frustrated magnets.\citep{Okubo2012,Leonov2015}
However, such phase exists only at external fields $B$ comparable
to the antiferromagnetic exchange $J$, which is extremely large in
most antiferromagnets. Our Monto Carlo simulations reveal that nonzero
topological charge takes place at low fields and elevated temperatures.
It can be thus easily accessible in experiments. We also proposed
a real frustrated system, the Mn-Bi mono-layer film, to enable thermally
driven topology. In this material, the DM interaction is exceedingly
large so that other topological non-trivial phases at the high field
are also discussed. 

\section{Method}

\subsection{Monto Carlo Simulations}

Numerical Metropolis Monte Carlo simulations were employed \citep{Metropolis_1949}
iteratively to generate a Markov chain \citep{Buhrandt_2013} of spin
configurations, which was then used to derive the thermal average
of the topological charge by employing the Berg formula \citep{Berg_1981}
with the method mentioned in Ref. \citep{Hou2017}. Periodic boundary
conditions were used for 2D hexagonal lattices with sizes $96\times96$
unless otherwise noted. Local magnetic moment $S$ is normalized so
that all the parameters $J$, $D$, $K_{u}$ and $B$ in the Hamilontian
as well as temperature $T$ have the dimensions of energy and $J$
is set as the units of both energy and temperature. Averages over
$2.56\times10^{6}$ ensembles are performed at each temperature step
during the annealing procedure which starts from $10.0J$, a very
high temperature. 

\subsection{First-principles calculations}

The spin-polarized first-principle calculations using the project
augmented wave pseudopotential (PAW) \citep{PAW_1994,Kresse_1999}
implemented in VASP package \citep{Kresse_1996_CMS,Kresse_1996_PRB}
were performed for calculating electronic and magnetic properties
of the Mn-Bi monolayer film. Local density approximation (LDA) \citep{Perdew1981}
was employed for the exchange-correlation functional. The wave functions
were expanded in plane waves with an energy cutoff of 600 eV throughout
calculations. The $k$ points were sampled on a $\Gamma$-centered
$15\times15$ mesh in the 2D Brillouin zone of unit cell containing
one A-B (Mn-Bi) site. Non-collinear magnetic calculations with SOC
are included in total energy calculations for obtaining the magnitude
of parameters $J$, $D$, $K_{u}$ in the Hamiltonian. 

\section{Results and Discussions}

\subsection{Topological charge driven by thermal fluctuations}

\begin{figure}
\includegraphics[width=1\columnwidth]{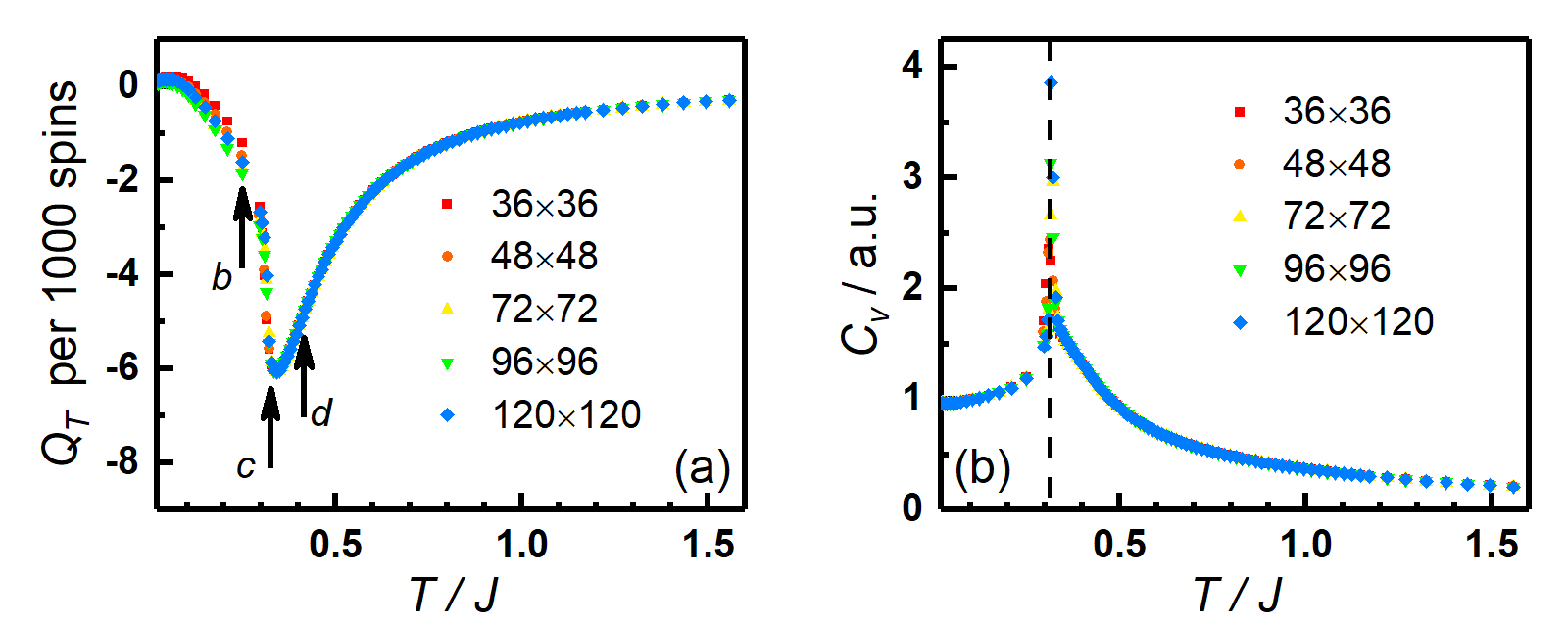}

\caption{(a) Topological charge density as a function of temperature in the
annealing process with various lattice sizes. (b) Temperature dependence
of specific heat with various lattice sizes. Three dashed lines labeled
$b$, $c$ and $d$ corresponding to $T=0.249J$, $0.328J$ and $0.401J$
respectively are for snapshots in Fig. \ref{fig:snapshots} }
\label{fig:first}
\end{figure}

We choose the parameters in the Hamiltonian with $D=0.40J$, $K_{u}=0.20J$
and $B=0.40J$. The relation between average topological charge $Q_{T}$
and temperature is shown in Fig. \ref{fig:first}. Topological charge
density is zero at very low temperature while it has a non-zero region
in finite temperature. The valley position is at $T=0.341J$, with
maximum density of $|Q_{T}$| about $6.07$ per 1000 spins. At very
high temperature, $Q_{T}$ again converges to zero due to the topological
triviality of a completely random phase. The same calculations were
performed for lattices with sizes ranging from $36\times36$ to $120\times120$.
Almost no difference could be found among them, indicating the immunity
to the finite-size effect. This robustness of the deep dip of the
topological charge should be the scaling-free atomic scale physics. 

The valley position of topological charges is close to the phase transition
temperature of spin-ordering. Specific heat $C_{v}$ as a function
of temperature is shown in Fig. \ref{fig:first}. A discontinuous
point appears at $T=0.315J$, indicating a second-order phase transition.
To further understand the relation between topological charge and
the phase transitions, we took snapshots of spin states at $T=0.020J$,
$0.249J$, $0.328J$ and $0.401J$, respectively. 

\begin{figure}
\includegraphics[width=1\columnwidth]{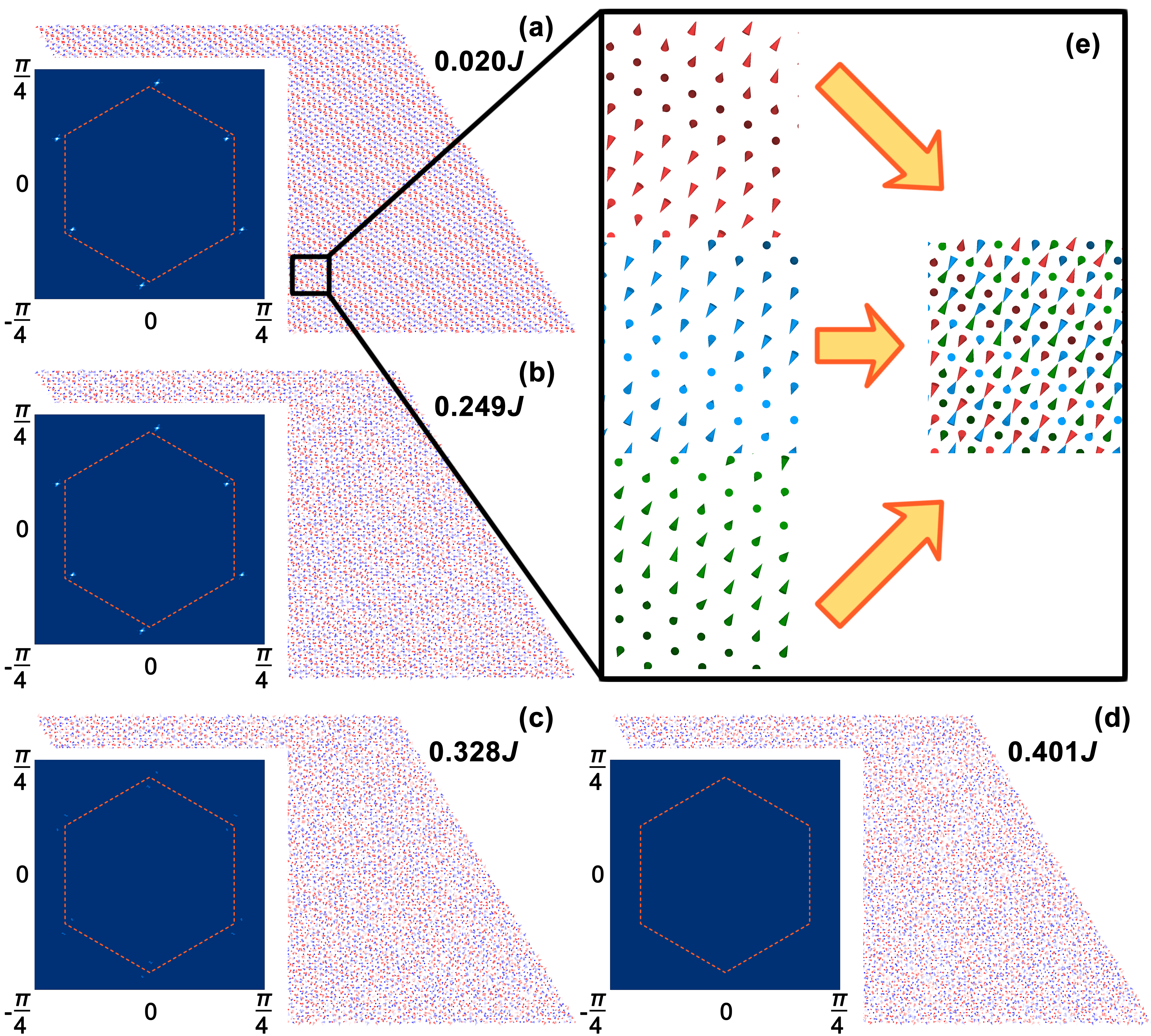}

\caption{Snapshots and corresponding reciprocal-space images by fast Fourier
transform(FFT) at (a) $T=0.020J$, (b) $T=0.249J$, (c)$T=0.328J$,
(d)$T=0.401J$ labeled as dashed line in Fig. \ref{fig:lattice_AB}(a).
(e) gives the zoom-in view of (a) to display the combination of three
sublattices with helix spin states. }
\label{fig:snapshots}
\end{figure}

The same spin-ordering phases take place at very low temperature $T=0.02J$
{[}Fig. \ref{fig:snapshots}(a){]} and low temperature $T=0.249J$
{[}Fig. \ref{fig:snapshots}(b){]}, according to both real space spin
textures and corresponding fast Fourier transformation (FFT) images.
Three neighbor spins in each triangle belongs to three distinct sublattices
in the hexagonal lattice, respectively. The second nearest spins are
in the same sublattice and tend to be parallel due to nearest neighbor
antiferromagnetic coupling. Then each sublattice forms one helical
state, shown in Fig. \ref{fig:snapshots}(e). The three sublattices
have the same $q$ vector so that it is called the single-$q$ state.
As a result, in the FFT images, six spots appear in the first Brillouin
zone, corresponding to the $\sqrt{3}\times\sqrt{3}$ spin-reconstruction
supercell and the wave vector of each helix corresponds to the offset
from each corner of the auxiliary red dashed hexagon in each FFT image
. It is obvious that this spin ordering has zero topological charge.
It is consistent with previous studies that in antiferromagnets\citep{Zhang_2016,Goebel2017}
or frustrated systems\citep{Okubo2012,Leonov2015}, no skyrmion or
other topological non-trivial spin textures can be found in the low
field region. 

At $T=0.328J$ {[}Fig. \ref{fig:snapshots}(c){]} which is just above
the phase transition temperature $0.315J$, the spin texture is disordered.
No spin-ordered texture can be found in real space, corresponding
to no spots found in the reciprocal space image. But at that temperature,
the topological charge density is $-5.91$ per 1000 spins, a definitely
non-zero value. The same situation happens at higher $T=0.401J$ {[}Fig.
\ref{fig:snapshots}(d){]}, which is higher than both the phase transition
temperature and valley position of topological charges. The topological
charge density is $-5.09$ per 1000 spins while the spin texture is
totally disordered. Thus, although the valley position of topological
charge is almost the phase transition temperature of spin-ordering,
the deep dip of topological charge density at finite temperature does
not correspond to any ordered phase because topological charge with
the definition in Eq. \ref{eq:tcharge} respects the rotational symmetry,
so that it is unable to serve as an order parameter which corresponds
to a spin-ordering phase.

It is a system which is topologically trivial at zero temperature
but has non-zero topological charge at finite temperature, including
temperatures above phase transition temperature. It has the same behavior
as the thermal-driven topology in chiral magnets\citep{Hou2017}.
According to the definition in Eq. \ref{eq:tcharge}, topological
charge $Q_{T}$ respects the spatial inversion symmetry but breaks
the time-reversal symmetry. The inversion symmetry bring about a quadratic
relation as the lowest order between $Q_{T}$ and magnetic field $D$,
which is spatial inversion odd; magnetic field $B$ breaks the time-reversal
symmetry so that $Q_{T}$ is proportional to $B$. Based on this,
the magnitude of topological charge relates to magnetic field and
DM interactions as: 
\begin{equation}
Q_{T}\propto D^{2}B.\label{eq:Q_DB}
\end{equation}

\begin{figure}
\includegraphics[width=1\columnwidth]{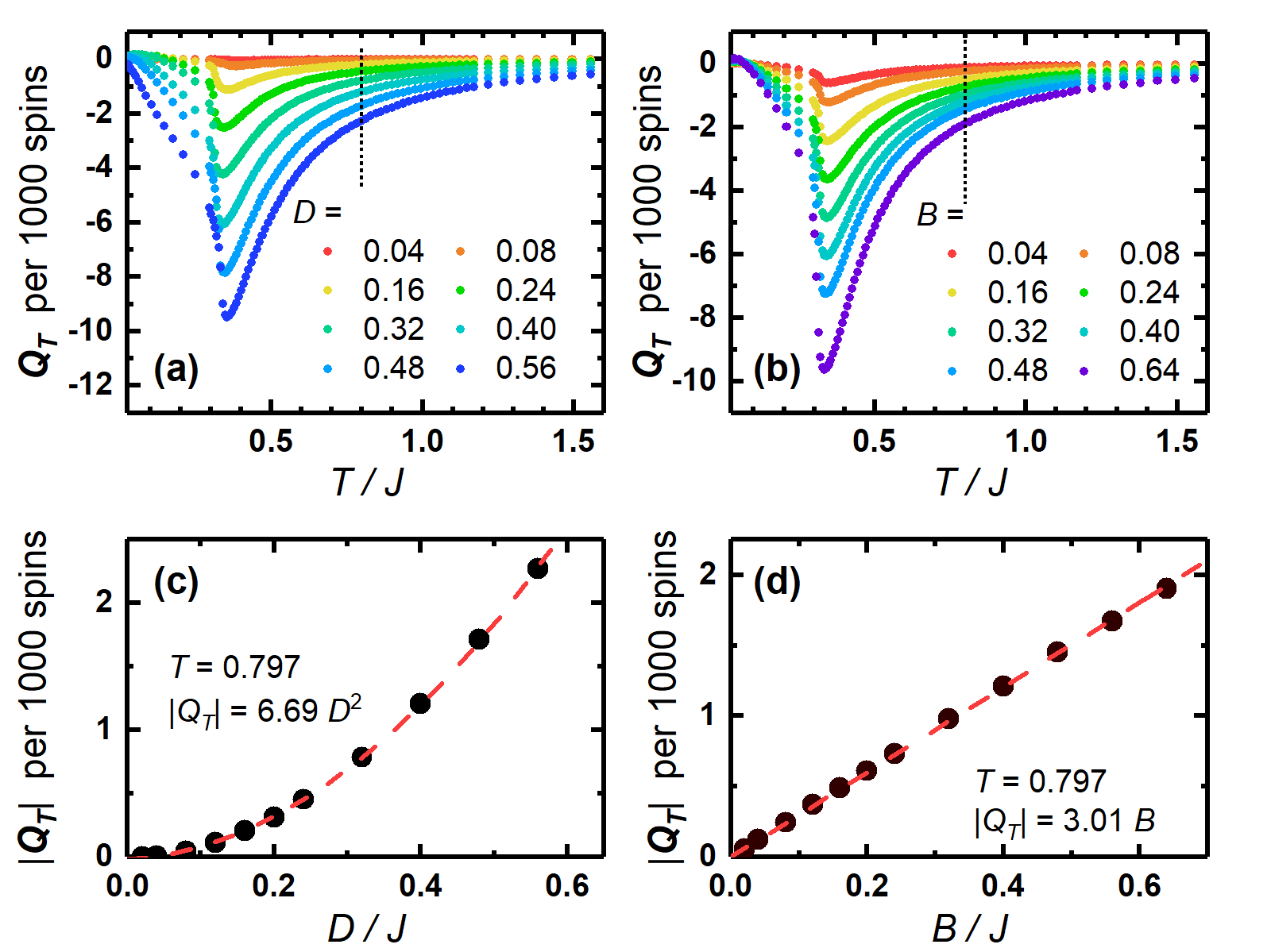}

\caption{Relationship between topological charge, and DM interaction as well
as external magnetic field. Topological charge density as a function
of temperature with various (a) $D$ and (b) $B$. In (a), $B=0.4J$
and $K_{u}=0.2J$ are fixed and in (b), $D=0.4J$ and $K_{u}=0.2J$
are fixed. The value of topological charge density $|Q_{T}|$ as a
function of (c) $D$ and (d) $B$ at $T=0.797J$ which is shown as
the dotted line in (a) and (b). }
\label{fig:chg_DB}
\end{figure}

The same argument applies to current system and a similar relation
is held. Fig. \ref{fig:chg_DB}(a)(c) shows the $D$ dependence of
topological charge density. Annealing curves show that the valley
positions of topological charge density are almost unchanged while
the valley value increases as $D$ increases. According to Fig. \ref{fig:chg_DB}(c),
the value of topological charge density $|Q_{T}|$ has a quadratic
relation with $D$ at $T=1.49J$. This quadratic relation is also
robust all the way to high temperatures. On the other hand, the relation
between topological charge density and $B$ is shown in Fig. \ref{fig:chg_DB}(b)(d).
An upward trend in $|Q_{T}|$ with the increasing $B$ is quite similar
with that of $|Q_{T}|-D$ relation in terms of annealing curves {[}Fig.
\ref{fig:chg_DB}(b){]}. The detailed investigation shows that $|Q_{T}|$
is proportional to $B$ {[}Fig. \ref{fig:chg_DB}(d){]}. 

Although the non-trivial topology in this frustrated system has exactly
the same relation with $D$ and $B$ as the thermally driven topology
in chiral magnets in high temperature, significant difference, however,
appears in low magnetic field and low temperature region. In chiral
magnets, skyrmions appear when both temperature and field is low,
so non-zero topological charge density can be found in this region.
Thermally driven topology is not the only mechanism causing non-trivial
topology.\citep{Hou2017} But for this spin frustrated system, here
one can only get the single-$q$ state with trivial topology below
phase transition temperature. The only source of non-zero topological
charge is the thermal average, so it is a system in which the magnetic
topological property is purely contributed by thermally driven topology. 

\begin{figure}
\includegraphics[width=1\columnwidth]{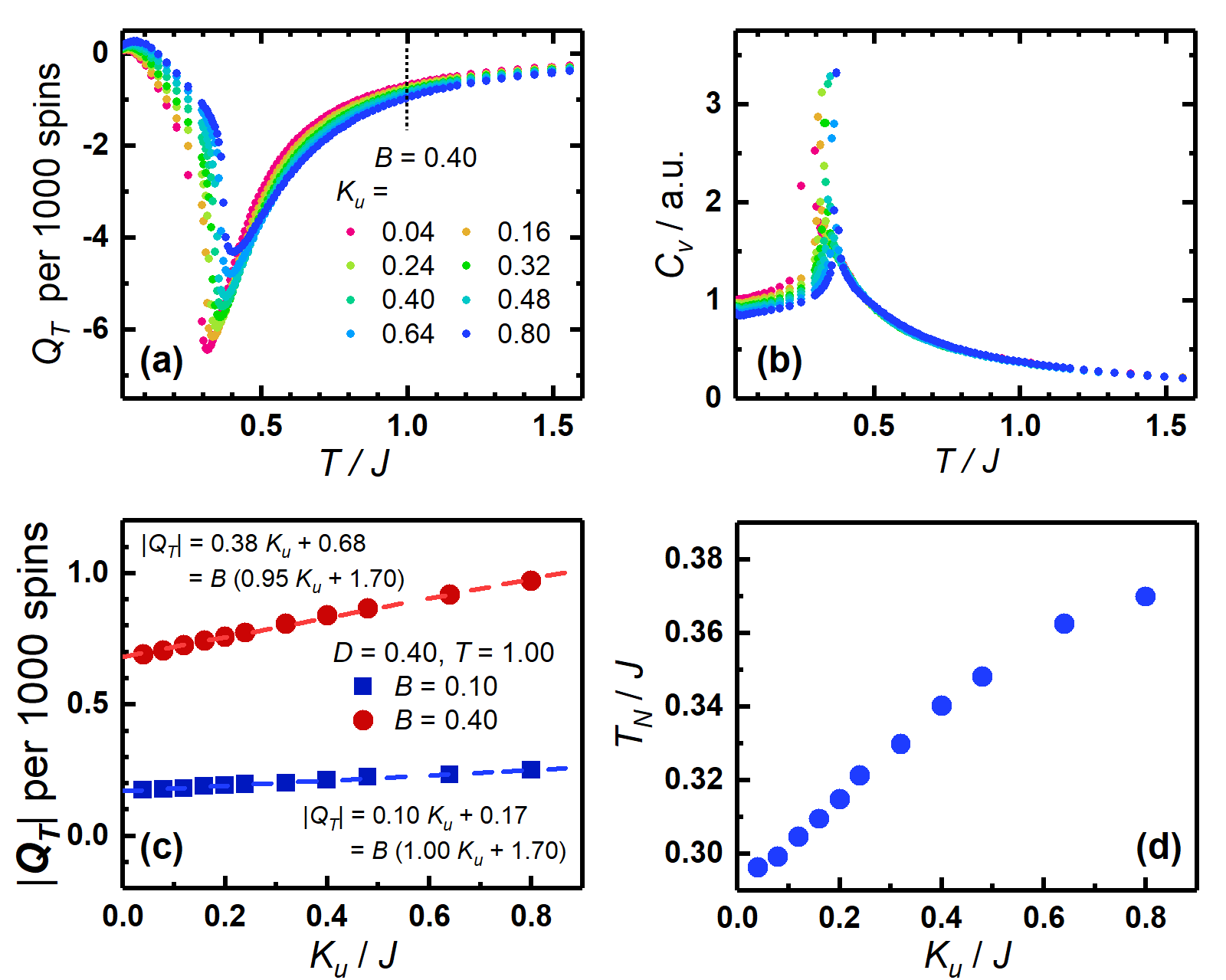}

\caption{Relationship between topological charge and uniaxial magnetic anisotropy.
(a) Topological charge density and (b) specific heat as functions
of temperature under various values of $K_{u}$ with other parameters
$D=0.40J$ and $B=0.40J$. (c) The value of topological charge density
as a function of $K_{u}$ at $T=1.00J$ (dotted line in (a)) with
two magnetic field values $B=0.10J$ and $0.40J$, respectively. (d)
Phase transition temperature $T_{N}$ as a function of $K_{u}$. }
\label{fig:chg_Ku}
\end{figure}

The uniaxial magnetic anisotropy $K_{u}$ dependence of topological
charge was also investigated. The results are shown in Fig. \ref{fig:chg_Ku}(a).
Other parameters were fixed at $D=0.40J$ and $B=0.40J$. As shown
in Fig. \ref{fig:chg_Ku}(a), the valley positions of topological
charges increase as $K_{u}$ is increasing, while the valley values
decrease. According to Fig. \ref{fig:chg_Ku}(c) with two magnetic
field values $B=0.10J$ and $0.40J$, the value of topological charge
has a linear relationship with $K_{u}$ in high temperature region.
This still obeys the symmetry of $|Q_{T}|$ since $K_{u}$ respects
both the inversion symmetry and the time-reversal symmetry. According
to the linear fitting parameters under two different magnetic field,
it can be written as $|Q_{T}|\propto B(K_{u}+1.70)$ or $B(1+\alpha K_{u})$
where $\alpha=0.59J^{-1}$.

$K_{u}$ has the similar behavior with external magnetic field at
high temperature since the mechanism of $K_{u}$ can be regarded as
an effective magnetic field $B_{eff,i}^{(K_{u})}=-\partial\left(\sum_{j}-\frac{1}{2}K_{u}S_{jz}^{2}\right)/\partial\mathbf{S}_{i}=K_{u}S_{iz}$
by the local moment itself applying on that moment. Since this effective
field is just an additional magnetic field, the intercept is non-zero
when $B\ne0$. On the other hand, the valley positions of topological
charges increase because the phase transition temperature is enlarged
by $K_{u}$. The $K_{u}$ dependence of specific heat $C_{v}$ as
a function of temperature is shown in Fig. \ref{fig:chg_Ku}(b). The
discontinuous point of $C_{v}$ move right as $K_{u}$ increases,
indicating the increaing of phase transition temperature. Fig. \ref{fig:chg_Ku}(d)
gives the relationship between phase transition temperature $T_{N}$
and $K_{u}$. The intercept value as $0.29J$ is the $T_{N}$ without
magnetic anisotropy and $T_{N}$ increases about $27\%$ when $K_{u}$
reaches $0.80J$.

As a result, based on Eq. \ref{eq:Q_DB}, the thermally-driven topological
charge in this system has the relationship
\begin{equation}
Q_{T}\propto D^{2}B(1+\alpha K_{u}),\alpha>0\label{eq:Q_DBK}
\end{equation}
in high temperature region. 

\subsection{Mn-Bi thin film from first principles calculations}

This 2D frustrated model is the prototype of many non-centrosymmetric
magnetic thin film systems. Here we chose manganese and bismuth as
the species of $A$ and $B$ sublattices in this model, respectively.
The hexagonal NiAs phase of MnBi is ferromagnetic with its high Curie
temperature over 600K and high coercivity with a rectangular hysteresis
loop due to large room-temperature perpendicular anisotropy\citep{Egashira1974,Guo1992,Di1992,Oppeneer1996,Yang2001}.
High Curie temperature is due to strong magnetization, as well as
strong exchange interaction, while large perpendicular room-temperature
anisotropy originates from strong spin-orbit coupling. Although there
is no DM interaction in bulk MnBi because the inversion symmetry is
respected, the inversion symmetry can be easily broken in an ultra-thin
film. An ideal result is the 2D film with the atomic scale.

\begin{figure}
\includegraphics[width=1\columnwidth]{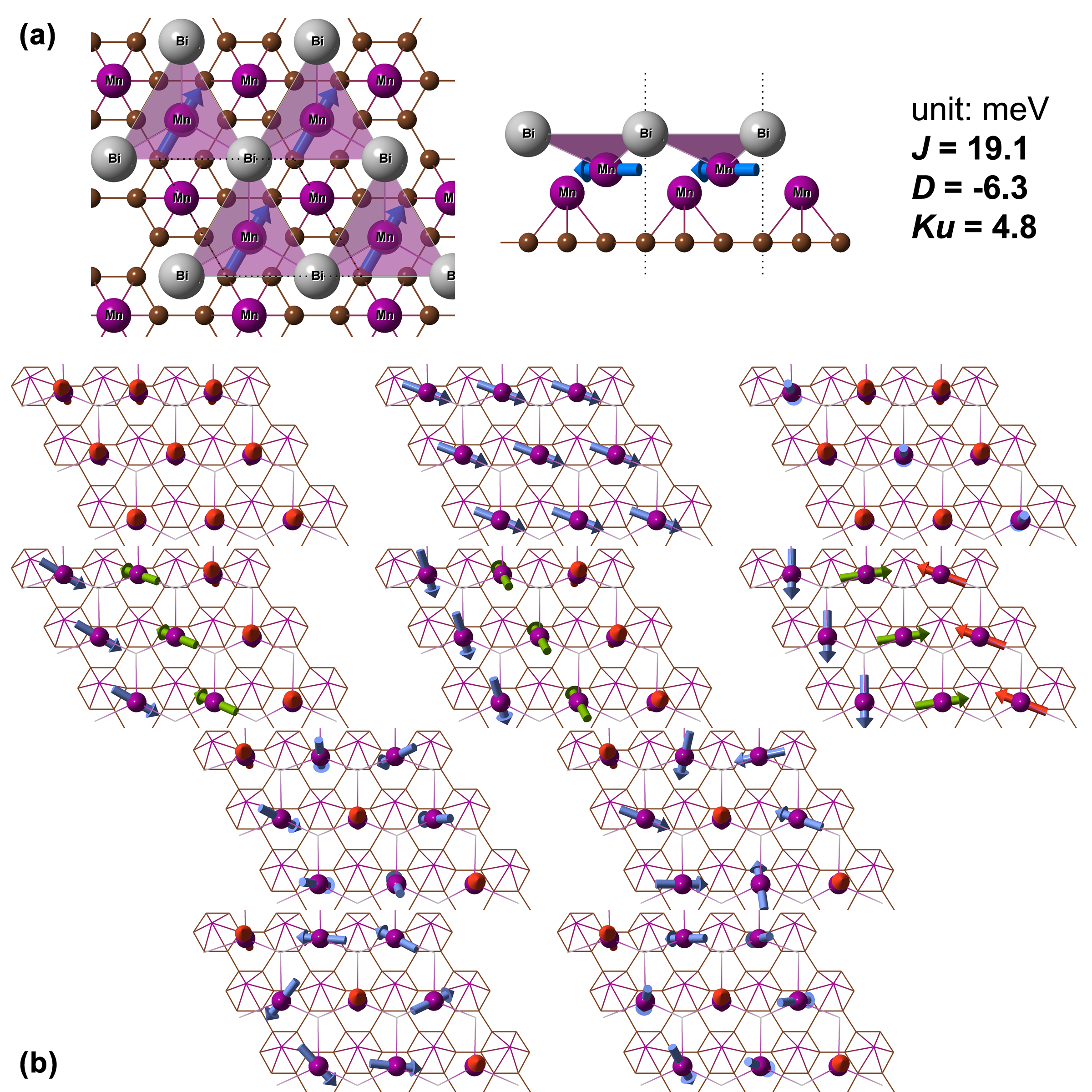}

\caption{(a) Top view and side view for the structure of 2 ML of Mn (violet
balls) and 1 ML of Bi (gray balls) on graphene (brown balls). (b)
Ten spin configurations used to build super rank equations for obtaining
$J$, $D$, $K_{u}$. The final result is also displayed. }
\label{fig:DFT}
\end{figure}

To construct Mn-Bi thin films, we chose a graphene sheet as the substrate
to stabilize the structure. This is because the in-plane lattice constant
of MnBi bulk of 4.28 � is quite close to that of $\sqrt{3}\times\sqrt{3}$
graphene lattice of 4.26 �. According to the LDA functional calculations,
the in-plane lattice constant is fixed at 4.24 � in one unit cell
(u.c.), which still matches the experimental results. Two mono-layers
(MLs) of Mn and one ML of Bi are placed on the graphene sheet. The
most stable structure with lowest total energy is shown in Fig. \ref{fig:DFT}(a).
Two Mn MLs are located between graphene and Bi ML so that the first
ML of Mn is located on the hexagonal hollow of graphene. According
to the result of the self-consistant calculations, The first ML of
Mn has no magnetization while the local magnetic momentum of the second
ML Mn is $3.2\mu_{B}$. 

To get the magnitude of the parameters $J$, $D$ and $K_{u}$, we
set ten non-collinear spin configurations on the second ML Mn atoms,
shown in Fig. \ref{fig:DFT}(b) and calculate the total energies for
each of them. A set of super rank equations based on the Hamiltonian
in Eq. \ref{eq:H_hex} was then built, and eventually all the parameters
were obtained. The result is $J=19.1$ meV/u.c., $D=-6.3$ meV/u.c.
and $K_{u}=4.8$ meV/u.c., also shown in Fig. \ref{fig:DFT}(a). $J$
is positive, indicating the antiferromagnetic coupling that was expected.
Due to the hexagonal lattice, this Mn-Bi thin film system is truly
a spin frustrated system. $K_{u}$ is also positive, so it is uniaxial
magnetic anisotropy. $D$ is negative due to the Bi ML located above
Mn MLs instead of below them, resulting in the clockwise direction
DM interaction around one magnetic moment. 

The magnitude of $D$ is quite large. This lattice model value of
$D$ can also be converted into the continuum model using $\tilde{D}=D\cdot a/\Omega$
where $a$ is the in-plane lattice constant and $\Omega$ is the volume
of the unit cell. Therefore, we have $\tilde{D}=-7.0\mathrm{mJ/m}^{2}$,
about four times larger than that in the chiral magnet FeGe \citep{Koretsune2015}.
The ratio of $D$ and $J$ is about 1/3, close to the parameters $D=0.4J$
in our Monto Carlo numerical simulations mentioned above. 

\subsection{Topology in the Mn-Bi thin film}

\begin{figure}
\includegraphics[width=1\columnwidth]{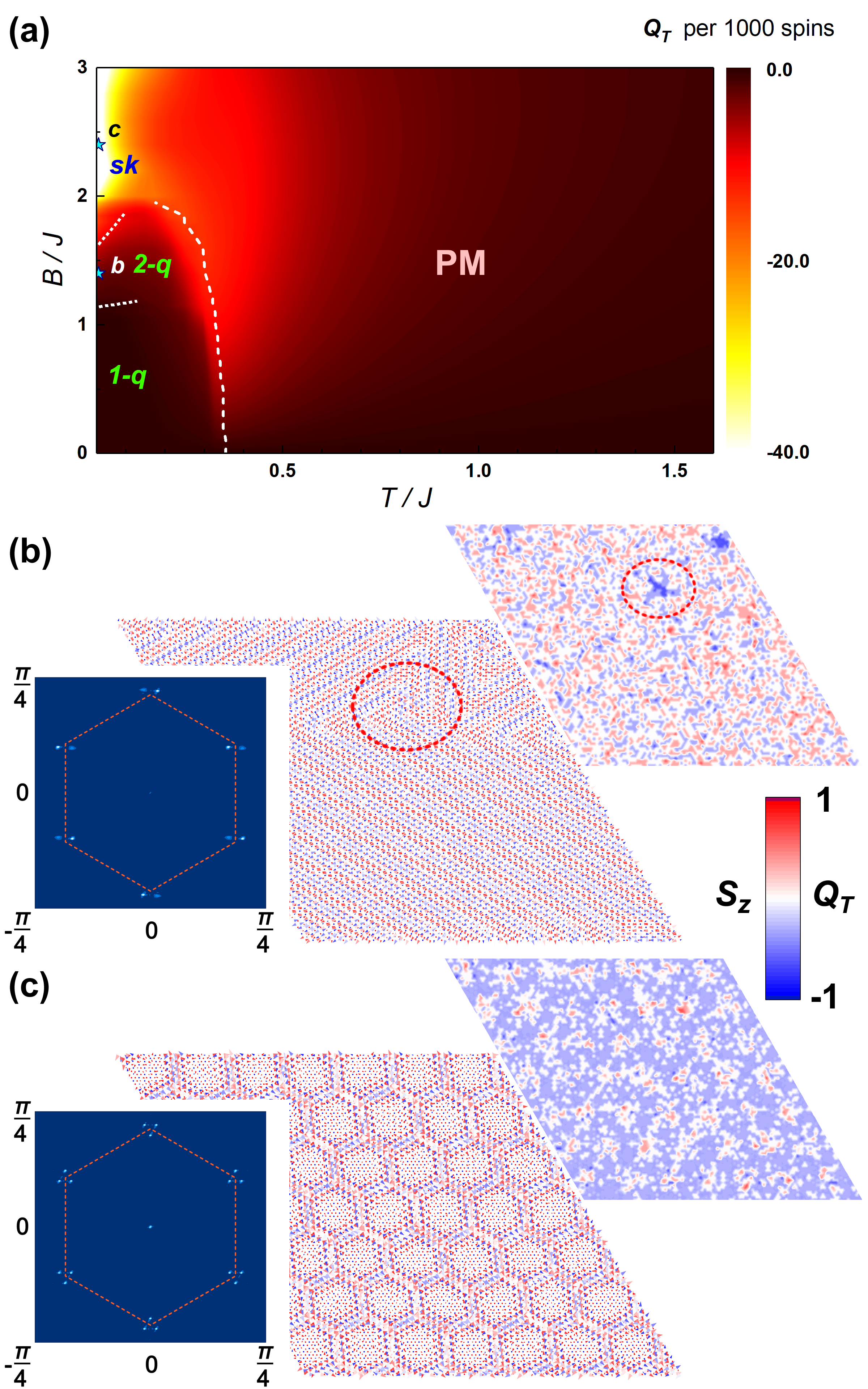}

\caption{Under Mn-Bi thin film system, (a) the phase diagram of topological
charge densities with the magnetic field and temperature dependence.
The valley positions are connected by a dashed line. Two white dotted
lines divides the region of single-$q$ (1-$q$), double-$q$ (2-$q$)
and skyrmion ($sk$) states, respectively. Star symbols labeled $b$
and $c$ correspond to the snapshots in (b) and (c). Snapshots of
spin textures and the corresponding FFT images and distribution of
topological charge density with (b) $B=1.4J$ and (c) $B=2.4J$ at
very low temperature $T=0.02J$. The color bar gives the colormap
of $z$ component of spins and the value of topological charge density.}
\label{fig:BT_phase}
\end{figure}

As all necessary parameters in Eq. \ref{eq:H_hex} were obtained in
the Mn-Bi thin film, we reformed Monto Carlo simulations to study
both the spin ordering and topology in this real system. Fig. \ref{fig:BT_phase}(a)
gives the $B-T$ diagram of topological charge density as a function
of the magnetic field $B$ and temperature $T$. At low magnetic field
region, topological charge density is zero at both very low and very
high temperature, while there is a ridge near the phase transition
temperature. At high temperature region, the absolute value of topological
charge densities increases as $B$ increases. Thus, it is consistsent
with the numerical simulations before, indicating that thermally driven
topology dominates the non-trivial topology at low field region in
the Mn-Bi thin film.

When $B>1.1J$, the value of topological charge becomes non-zero at
low temperature. According to the snapshot at $B=1.4J$ shown in Fig.
\ref{fig:BT_phase}(b), two main helical waves, namely double-$q$,
appear in the real space. The two helices form an angle of 120 degrees,
resulting in two spots near each corner of the auxiliary hexagon in
the reciprocal space. The density of topological charge also shows
that although the contribution of topological charge density looks
random in the whole lattice, the intersection between two helical
waves bring about the considerable negative numbers of topological
charges, shown in the dashed circles in Fig. \ref{fig:BT_phase}(b). 

As the magnetic field continues increasing, the number of topological
charge grows as well, especially at low temperatures. Around $B=1.8J$,
however, another phase transition takes place. The number of topological
charge increase swiftly and has the valley value of over $-58$ per
1000 atoms at $B=2.2J$. The snapshot at $B=2.4J$ gives an honeycomb-like
lattice of spin texture, shown in Fig. \ref{fig:BT_phase}(c). In
the reciprocal space, three spots appear near each corner of the auxiliary
hexagon so that there are three helical waves, of which the wave vector
has an angle of 120 degrees with other two, in each of the three sublattices.
The combination of three helical waves is nothing but the skyrmion
crystal lattice, so that each sublattice forms one skyrmion lattice.
The total spin texture is the combination of these three skyrmion
lattices and the centers of skyrmions together form one hexagonal
lattice. 

In most skyrmion lattices found in ferromagnets, each skyrmion gives
$-1$ topological charge. Then only 50 skyrmions in the $96\times96$
lattice, namely about$-5$ topological charge per 1000 spins, are
identified according to the spin texture in Fig. \ref{fig:BT_phase}(c).
It is far from the total value of $-55$ that the lattice really has.
It is because each three neighbor spins forming a triangle are belong
to different sublattices. Because of nearest neighbor antiferromagnetic
coupling, a large solid angle is formed among them, leading to the
large magnitude of the density of topological charge. So that the
density of topological charge, shown in \ref{fig:BT_phase}(c) gives
the almost uniformed density with negative value in whole lattice.
Thus, the skyrmion phase in this frustrated system has distinct non-trivial
topology with skyrmion phase in ferromagnets. Both double-$q$ and
skyrmion crystal phase at low temperatures consists with previous
studies on frustrated systems\citep{Okubo2012,Leonov2015}.

One should note that the critical magnetic field for skyrmion phase
is about $2.0J$. Since $J\sim20$ meV in the Mn-Bi thin film, the
critical field cannot be reached. Furthermore, at least 22 meV of
magnetic field, corresponding to $B=1.1J$, still a giant value, is
required to get the double-$q$ state. As a result, under a very wide
range of magnetic field strengths, no non-trivial topological phase
except for thermally driven topology can be detected in this spin
frustrated system, particularly in this Mn-Bi thin film system. 

\section{Summary}

In conclusion, we have studied the topological properties in a frustrated
hexagonal lattice. In a very wide range of low magnetic field, thermally
driven topology is the only non-trivial topology under finite temperature.
The magnitude of topological charge has the relationship $Q_{T}\sim D^{2}B(1+\alpha K_{u})$
with DM interaction $D$, magnetic field $B$ and uniaxial magnetic
anisotropy $K_{u}$ in high temperature region. The Mn-Bi thin film
system on graphene sheet is such a frustrated system with significant
DM interaction. The magnitude of$J$, $D$, and $K_{u}$ in Eq. \ref{eq:H_hex}
is identified via first principles calculations. Based on these parameters,
a $B-T$ phase diagram of topological charge is obtained. Although
both double-$q$ and skyrmion states are found at very high magnetic
field region, thermally driven topology is the only non-trivial topological
phase under a very wide field range. This makes it easier to be detected
by experiments of thermal magnon Hall effect.

This work was supported by the U.S. Department of Energy (DOE), Office of Science, Basic Energy Sciences (BES) under Award No. DE-SC0016424 and used the Extreme Science and Engineering Discovery Environment (XSEDE) under Grant No. TG-PHY170023 for first-principles calculations.

\end{document}